\documentclass[prd,preprint,groupaddress,showpacs,nofootinbib,tightenlines,amsmath]{revtex4}
\usepackage{graphics}
\usepackage{epsfig}

\def\bra#1{\ensuremath{\langle{#1}\vert}}
\def\ket#1{\ensuremath{\vert{#1}\rangle}}

\begin{document}
\title{Threshold $\pi^0$ photoproduction in relativistic chiral perturbation
theory
\thanks{Supported by the Deutsche Forschungsgemeinschaft through the
SFB~443}}

\author{M. Hilt, S. Scherer, and L. Tiator}

\affiliation{PRISMA Cluster of Excellence,
Institut f\"ur Kernphysik, Johannes-Gutenberg Universit\"at Mainz, D-55099 Mainz,
Germany
}
\date{January 23, 2013}
\preprint{MITP/13-008}
\begin{abstract}
   We present a calculation of $\pi^0$ photoproduction on the proton in
manifestly Lorentz-invariant baryon chiral perturbation theory up to and
including chiral order $q^4$.
   With the results we analyze the latest $\pi^0$ photoproduction data in the
threshold region obtained at the Mainz Microtron.
   In the calculation of observables and the fit of the low-energy constants,
we take $S$, $P$, and $D$ waves into account.
   We compare the results for the multipoles with the corresponding
single-energy analysis.
   Furthermore, we also fit the $O(q^4)$ heavy-baryon chiral
perturbation theory calculation and compare both results.
   We provide predictions for several polarization observables for future
experiments.
   Finally, we discuss the $\beta$ parameter of the unitarity cusp
which is related to the breaking of isospin symmetry.
\end{abstract}
\maketitle

\section{Introduction}

   Quantum chromodynamics (QCD), the gauge theory of the strong interactions, possesses an approximate global
symmetry due to the small masses of the $u$ and $d$ quarks.
   Exploiting this so-called $\textnormal{SU(2)}_\textnormal{L}\times \textnormal{SU}(2)_\textnormal{R}$
chiral symmetry and its spontaneous breaking to $\textnormal{SU}(2)_\textnormal{V}$, model-independent
predictions can be made for physical observables.
   Such a prediction is called a low-energy theorem (LET) and, in the case of pion photoproduction,
the first derivation of a LET for the charged channels by Kroll and Ruderman \cite{Kroll:1953vq} was
based on electromagnetic current conservation.
   For neutral pion photoproduction, the first sufficiently precise experimental data
\cite{Mazzucato:1986dz,Beck:1990da} for the $S$-wave electric dipole amplitude $E_{0+}$ \cite{Chew:1957tf}
showed a serious disagreement with the prediction based on current algebra and the partially-conserved
axial-vector current hypothesis \cite{DeBaenst:1971hp,Vainshtein:1972ih}.
   In Refs.~\cite{Bernard:1991rt,Bernard:1992nc}, the discrepancy between experiment and theory was
addressed with the aid of chiral perturbation theory (ChPT) which is an effective field theory of
QCD at low energies based on chiral symmetry (see, e.g., Refs.\ \cite{Scherer:2009bt,Scherer:2012zzd}
for an introduction).
   In particular, it was shown that $E_{0+}$ gets modified by certain nonanalytic loop contributions.
   Using heavy-baryon ChPT (HBChPT) \cite{Jenkins:1990jv,Bernard:1992qa}, Bernard \textit{et al.}
analyzed neutral pion photoproduction in Refs.\ \cite{Bernard:1994gm,Bernard:1995cj,Bernard:2001gz}.
   One motivation for introducing HBChPT was the fact that manifestly Lorentz-invariant (or relativistic)
ChPT (RChPT) seemingly had a problem concerning power counting when loops containing internal nucleon
lines come into play.
   In this case, a diagram apparently has contributions which are of lower order than determined
by the power counting.
   By choosing appropriate renormalization conditions, this problem was solved in
Refs.\ \cite{Becher:1999he,Fuchs:2003qc}.
   The crucial difference between HBChPT and RChPT is that, at a given order, the latter also includes
an infinite number of higher-order corrections.
   These corrections can be important as, e.g., in the case of the scalar nucleon form factor where one even
gets the wrong analytic behavior in HBChPT \cite{Becher:1999he}.
   Another example is the Fubini-Furlan-Rossetti sum rule \cite{Fub65}.
   The expansion around the nucleon mass shifts the pole positions in the $s$ and $u$ channels away
from the physical pole positions \cite{Pasquini:2004nq}.
    A further, non-perturbative approach for studying pion production beyond the threshold region
in a covariant way was developed in Ref.\ \cite{Gasparyan:2010xz}.
   The method makes use of the chiral effective Lagrangian up to and including $O(q^3)$ and
is based on the implementation of causality, coupled-channel unitarity, and electromagnetic
gauge invariance.

   In pion photoproduction, some of the higher-order corrections turn out to be large.
   While in Ref.\ \cite{Bernard:1994gm} an $O(q^3)$ calculation obtained LETs for the
$P$-wave amplitudes $P_1$ and $P_2$, the calculation at $O(q^4)$ \cite{Bernard:2001gz} gave
large corrections to these LETs.
   As one can see from the numerical values of the low-energy constants (LECs), there are still
important higher-order contributions missing.
   Here, we present a full one-loop  $O(q^4)$ RChPT calculation.
   We find that some of these higher-order contributions are included in the relativistic case.
   In addition, we also analyze $D$ waves and show that at order $O(q^4)$ there is another
LEC, which mainly affects the $E_{2-}$ multipole and, through mixing, also $E_{0+}$.
   We will focus on the latest data obtained at the Mainz Microtron (MAMI) \cite{Hornidge:2012ca}
which give very precise results for the differential cross section and the photon asymmetry $\Sigma$
in the threshold region and which are, therefore, well-suited to pin down the LECs and, with that,
the multipoles.

   This work is organized as follows.
   In Sec.\ II we present the formalism for neutral pion photoproduction.
   Section III gives a short introduction into the framework of ChPT.
   In Sec.\ IV we present and discuss our results.
   We give predictions for several polarization observables and compare them with the predictions
of the Dubna-Mainz-Taipei (DMT) model \cite{Kamalov:2000en}.
   We also show a fit of the HBChPT results to the new data to have a better comparison with RChPT.
   Finally, we analyze the so-called $\beta$ parameter of the unitarity cusp
\cite{Faldt:1979fs,Bernstein:1998ip}.
   Section V closes with a short summary.
   Some technical details of our calculation can be found in the appendices.

\section{Matrix element}
\begin{figure}[htbp]
    \centering
        \includegraphics[width=0.5\textwidth]{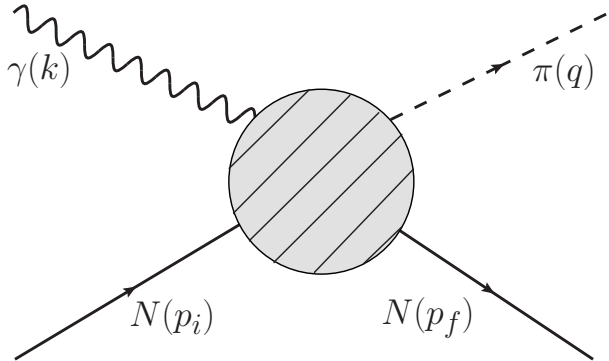}
    \caption{Kinematics of pion photoproduction. The solid line represents the nucleon with incoming
    momentum $p_i$ and outgoing momentum $p_f$, respectively. The wiggly line represents the photon
    with momentum $k$ and the dashed line denotes the pion with momentum $q$. The blob symbolically
    stands for all contributions to the process.}
    \label{fig:process}
\end{figure}

   Pion photoproduction on the nucleon is the creation of a pion from a nucleon via the absorption of a photon,
\begin{equation}
N+\gamma\rightarrow N'+\pi,
\end{equation}
where $N$ ($N'$) denotes the nucleon in the initial (final) state, $\gamma$ represents the photon,
and $\pi$ symbolically stands for the appropriate pion.
   Here, we only discuss the production of a $\pi^0$ on the proton.
   The kinematics of the process is depicted in Fig. \ref{fig:process}.
   The usual Mandelstam variables $s$, $t$, and $u$ are defined as
\begin{equation}
s=(p_i+k)^2=(p_f+q)^2,\ u=(p_i-q)^2=(p_f-k)^2, \ t=(p_i-p_f)^2=(q-k)^2,\label{eqn:mandelstam}
\end{equation}
and fulfill
\begin{equation}
s+t+u=2m_N^2+M_\pi^2,
\end{equation}
where $m_N$ and $M_\pi$ denote the nucleon mass and the pion mass, respectively.
   In the center-of-mass (cm) frame, the energies of the photon, $E_\gamma$, and the pion, $E_\pi$,
are given by
\begin{equation}
E_\gamma=\frac{W^2-m_N^2}{2W},\ E_\pi=\frac{W^2+M_\pi^2-m_N^2}{2W},
\end{equation}
where $W=\sqrt{s}$ is the cm total energy.
  In the lab frame, the photon energy $E_\gamma^\textnormal{lab}$ is given by
\begin{equation}
E_\gamma^\textnormal{lab}=\frac{W^2-m_N^2}{2m_N}.
\end{equation}
   The cm scattering angle $\Theta_\pi$ between the pion three-momentum and the $z$-axis,
defined by the incoming photon, can be related to the Mandelstam variable $t$ via
\begin{equation}
t=M_\pi^2-2(E_\gamma E_\pi-|\vec{k}||\vec{q}\hspace{0.05cm}|\cos\Theta_\pi).
\end{equation}

   For linearly polarized photons, the differential cross section
$\frac{d\sigma}{d\Omega}$ in the cm frame can be written as
\begin{equation}
\frac{d\sigma}{d\Omega}(\Theta_\pi,\phi)=\sigma_0(\Theta_\pi)
\left[1-\Sigma(\Theta_\pi)\cos 2\phi\right],
\end{equation}
where $\sigma_0(\Theta_\pi)$ is the unpolarized cross section and $\Sigma(\Theta_\pi)$ denotes
the photon asymmetry.
   The azimuthal angle $\phi$ is defined as the angle between the polarization vector
of the photon and the reaction plane spanned by the nucleon and pion three-momenta.

   Using the so-called Ball amplitudes \cite{Ball:1961zza}, the matrix element of pion
photoproduction can be parametrized in a Lorentz-covariant way,
\begin{equation}
-ie\epsilon_\mu\bra{N'\pi}J^\mu(0)\ket{N}=\epsilon_\mu\bar{u}(p_f)\left(\sum_{i=1}^8 B_i V_i^\mu\right) u(p_i).
\label{eqn:ball}
\end{equation}
   In Eq.\ (\ref{eqn:ball}), $\epsilon_\mu$ denotes the polarization vector of the photon,
$J^\mu$ is the electromagnetic current operator in units of the elementary charge $e>0$,
and $u(p_i)$ and $\bar{u}(p_f)$ are the Dirac spinors of the nucleon in the initial and
final states, respectively.
   In the following, our convention differs slightly from Ball's original definition.
   We use
\begin{align}
&V_1^\mu=\gamma^\mu\gamma_5, & &V_2^\mu=\gamma_5 P^\mu,\nonumber\\
&V_3^\mu=\gamma_5 q^\mu, & &V_4^\mu=\gamma_5 k^\mu, \nonumber\\
&V_5^\mu=\gamma^\mu k_\nu\gamma^\nu\gamma_5, & &V_6^\mu=k_\nu\gamma^\nu\gamma_5 P^\mu,\nonumber\\
&V_7^\mu=k_\nu\gamma^\nu\gamma_5 q^\mu, & &V_8^\mu=k_\nu\gamma^\nu\gamma_5 k^\mu,
\end{align}
with $P=(p_i+p_f)/2$.
   Only six of the amplitudes survive in pion photoproduction, as $\epsilon\cdot k=0$.
   The remaining amplitudes are related because of current conservation,
\begin{equation}
B_1+B_6 k\cdot P+B_7 k\cdot q=0,\qquad B_2 k\cdot P+B_3 k\cdot q=0,
\end{equation}
so one ends up with only four independent structures.

   In the threshold region, pion photoproduction is commonly analyzed in terms of a multipole decomposition.
   The usual definition of the matrix element in the cm frame \cite{Chew:1957tf} is related to Eq.\ (\ref{eqn:ball})
via
\begin{equation}
\epsilon_\mu\bar{u}(p_f)\left(\sum_{i=1}^8 B_i V_i^\mu\right) u(p_i)=\frac{4\pi W}{m_N}\chi_f^\dagger\mathcal{F}\chi_i,
\end{equation}
where $\chi_i$ and $\chi_f$ denote initial and final Pauli spinors.
   In the Coulomb gauge ($\epsilon_0=0$, $\vec{k}\cdot\vec\epsilon=0$), $\mathcal{F}$ may be written as follows,
\begin{eqnarray}
\mathcal{F}&=&i \vec{\sigma}\cdot\vec{\epsilon}\mathcal{F}_1
+\vec{\sigma}\cdot\hat{q}\ \vec{\sigma}\cdot(\hat{k}\times\vec{\epsilon})\mathcal{F}_2
+i\vec{\sigma}\cdot\hat{k}\ \hat{q}\cdot\vec{\epsilon}\mathcal{F}_3
+i \vec{\sigma}\cdot\hat{q}\ \hat{q}\cdot\vec{\epsilon}\mathcal{F}_4,
\end{eqnarray}
where $\hat{q}$ and $\hat{k}$ denote unit vectors in the direction of the pion and the
photon, respectively.
   Introducing $x=\cos\Theta_\pi=\hat{q}\cdot\hat{k}$,
the so-called Chew-Goldberger-Low-Nambu (CGLN) amplitudes $\mathcal{F}_i$ can be expressed
in terms of energy-dependent multipole amplitudes as
\begin{align}
\mathcal{F}_1=&\sum_{l=0}^\infty\Big\{\big[lM_{l+}+E_{l+}\big]P'_{l+1}(x)
+\big[(l+1)M_{l-}+E_{l-}\big]P'_{l-1}(x)\Big\},\nonumber\\
\mathcal{F}_2=&\sum_{l=1}^\infty\Big\{(l+1)M_{l+}+lM_{l-}\Big\}P'_l(x),\nonumber\\
\mathcal{F}_3=&\sum_{l=1}^\infty\Big\{\big[E_{l+}-M_{l+}\big]P''_{l+1}(x)
+\big[E_{l-}+M_{l-}\big]P''_{l-1}(x)\Big\},\nonumber\\
\mathcal{F}_4=&\sum_{l=2}^\infty\Big\{M_{l+}-E_{l+}-M_{l-}-E_{l-}\Big\}P''_l(x).\label{eqn:multdecomp}
\end{align}
   In Eq.\ (\ref{eqn:multdecomp}), $P_l(x)$ is a Legendre polynomial of degree $l$,
$P'_l=dP_l/dx$ and so on, with $l$ denoting the orbital angular momentum
of the pion-nucleon system in the final state.
   The multipoles $E_{l\pm}$ and $M_{l\pm}$ refer to transitions caused by electric and magnetic radiation,
respectively, and the subscript $l\pm$ denotes the total angular momentum $j=l\pm1/2$ in the final state.
   These multipoles have several advantages.
   First of all, one can match their quantum numbers to those of nucleon resonances in order
to analyze the excitation spectrum.
   Furthermore, while the amplitudes $\mathcal{F}_i$ depend on the cm angle of the reaction in a
complicated manner, this dependence can be completely projected out in the case of the multipoles,
as the Legendre polynomials form an orthogonal basis.
   The results read \cite{Davidson:1995jm}
\begin{align}
E_{l+}=&\int_{-1}^1\frac{dx}{2(l+1)}\Big[P_l \mathcal{F}_1-P_{l+1}\mathcal{F}_2\nonumber\\
&+\frac{l}{2l+1}(P_{l-1}-P_{l+1})\mathcal{F}_3+\frac{l+1}{2l+3}(P_l-P_{l+2})\mathcal{F}_4\Big],\nonumber\\
E_{l-}=&\int_{-1}^1\frac{dx}{2l}\Big[P_l\mathcal{F}_1-P_{l-1}\mathcal{F}_2\nonumber\\
&-\frac{l+1}{2l+1}(P_{l-1}-P_{l+1})\mathcal{F}_3+\frac{l}{2l-1}(P_l-P_{l-2})\mathcal{F}_4\Big],\nonumber\\
M_{l+}=&\int_{-1}^1\frac{dx}{2(l+1)}\left[P_l\mathcal{F}_1-P_{l+1}\mathcal{F}_2
-\frac{1}{2l+1}(P_{l-1}-P_{l+1})\mathcal{F}_3\right],\nonumber\\
M_{l-}=&\int_{-1}^1\frac{dx}{2l}\left[-P_l\mathcal{F}_1+P_{l-1}\mathcal{F}_2
+\frac{1}{2l+1}(P_{l-1}-P_{l+1})\mathcal{F}_3\right].\label{eqn:multipoles}
\end{align}
   In addition, in the threshold region one needs only few multipoles  to describe physical observables.
   Traditionally, only $S$ and $P$ waves were used, resulting in the following simple expressions for the
unpolarized differential cross section $\sigma_0$ and the photon asymmetry $\Sigma$ \cite{Drechsel:1992pn},
\begin{align}
\sigma_0(\Theta_\pi)&=\frac{|\vec{q}|}{|\vec{k}|}\left(A+B \cos\Theta_\pi+C \cos^2\Theta_\pi\right),\\
\Sigma(\Theta_\pi)&=
\frac{|\vec{q}|\sin^2\Theta_\pi}{2|\vec{k}|\sigma_0(\Theta_\pi)}\left(|P_3|^2-|P_2|^2\right),\label{eqn:csandasym}
\end{align}
where
\begin{align}
A&=|E_{0+}|^2+\frac{1}{2}(|P_2|^2+|P_3|^2),\nonumber\\
B&=2\textnormal{Re}(E_{0+}P_1^*),\nonumber\\
C&=|P_1|^2-\frac{1}{2}(|P_2|^2+|P_3|^2).
\end{align}
   The above-mentioned linear combinations of $P$ waves, $P_i$, are defined as
\begin{align}
P_1&=3E_{1+}+M_{1+}-M_{1-},\nonumber\\
P_2&=3E_{1+}-M_{1+}+M_{1-},\nonumber\\
P_3&=2M_{1+}+M_{1-}.
\label{eqn:Pi}
\end{align}
   Furthermore, we show predictions for polarization observables, namely, target asymmetry $T$,
recoil polarization $P$, and beam-target asymmetries $E$, $F$, $G$, and $H$ \cite{Drechsel:1992pn},
\begin{align}
T(\Theta_\pi)=&\frac{|\vec{q}|\sin\Theta_\pi}{|\vec{k}|\sigma_0(\Theta_\pi)}
\left\{\textnormal{Im}[E_{0+}^*(P_2-P_3)]+\cos\Theta_\pi\textnormal{Im}[P_1^*(P_2-P_3)]\right\},\nonumber\\
P(\Theta_\pi)=&-\frac{|\vec{q}|\sin\Theta_\pi}{|\vec{k}|\sigma_0(\Theta_\pi)}
\left\{\textnormal{Im}[E_{0+}^*(P_2+P_3)]+\cos\Theta_\pi\textnormal{Im}[P_1^*(P_2+P_3)]\right\},\nonumber\\
E(\Theta_\pi)=&\frac{|\vec{q}|}{|\vec{k}|\sigma_0(\Theta_\pi)}
\left\{|E_{0+}|^2+\textnormal{Re}(P_3P_2^*)+2\cos\Theta_\pi\textnormal{Re}(P_1E_{0+}^*)\nonumber\right.\\
&\left.+\cos^2\Theta_\pi[|P_1|^2-\textnormal{Re}(P_3P_2^*)]\right\},\nonumber\\
F(\Theta_\pi)=&\frac{|\vec{q}|\sin\Theta_\pi}{|\vec{k}|\sigma_0(\Theta_\pi)}
\left\{\textnormal{Re}[E_{0+}^*(P_2-P_3)]+\cos\Theta_\pi\textnormal{Re}[P_1^*(P_2-P_3)]\right\},\nonumber\\
G(\Theta_\pi)=&-\frac{|\vec{q}|\sin^2\Theta_\pi}{|\vec{k}|\sigma_0(\Theta_\pi)}
\textnormal{Im}(P_3P_2^*),\nonumber\\
H(\Theta_\pi)=&\frac{|\vec{q}|\sin\Theta_\pi}{|\vec{k}|\sigma_0(\Theta_\pi)}
\left\{\textnormal{Im}[E_{0+}^*(P_2+P_3)]+\cos\Theta_\pi\textnormal{Im}[P_1^*(P_2+P_3)]\right\}.
\end{align}
   In Refs.\ \cite{FernandezRamirez:2009su,FernandezRamirez:2009jb}, the importance of $D$ waves
was pointed out, especially of the $E_{2-}$ multipole.
   Even though their numerical values are small, they strongly affect the extraction of
other multipoles through interference with large $P$ waves.
   The relevant formulas for the observables are rather lengthy when the $D$ waves are included,
so we refer to Ref.\ \cite{FernandezRamirez:2009jb} for further details.
   Usually, in the threshold region the $D$ waves are assumed to be given entirely by Born contributions.
   In the next section, a low-energy constant is discussed which explicitly influences
$E_{2-}$ and, because this multipole strongly mixes with $E_{0+}$,
can change the determination of $E_{0+}$ significantly.
   In the threshold region, the multipoles $\mathcal{M}_{l\pm}$ ($\mathcal{M}=E,M$) are proportional
to $|\vec{q}|^l$.
   To get rid of this purely kinematical dependence, one introduces reduced multipoles
$\overline{\mathcal{M}}_{l\pm}$ via
\begin{equation}
\overline{\mathcal{M}}_{l\pm}=\frac{\mathcal{M}_{l\pm}}{|\vec{q}|^l}.
\end{equation}

   In the isospin-symmetric case, the amplitude for producing a pion with Cartesian isospin index
$a$ can be decomposed as
\begin{equation}
M(\pi^a)=\chi_f^\dagger(i\epsilon^{a3b}\tau^bM^{(-)}+\tau^aM^{(0)}+\delta^{a3}M^{(+)})\chi_i\label{eqn:isospin},
\end{equation}
where $\chi_i$ and $\chi_f$ denote the isospinors of the initial and final nucleons, respectively,
and $\tau^a$ are the Pauli matrices.
   Here, the only relevant physical channel is given by
\begin{equation}
M(\gamma+p\rightarrow\pi^0+p)=M^{(0)}+M^{(+)}.
\end{equation}
   This will be important when it comes to the determination of the LECs in the next section.

\section{Theoretical Framework}
   Chiral perturbation theory is the low-energy effective field theory of QCD based on an approximate
chiral symmetry and its spontaneous breakdown \cite{Weinberg:1978kz,Gasser:1983yg}.
   In the one-nucleon sector, pions and nucleons are used as the effective degrees of freedom \cite{Gasser:1987rb}.
   Starting from the most general Lagrangian in combination with a suitable power-counting scheme,
observables are calculated in a momentum and quark-mass expansion.
   At first, the correspondence between the chiral and loop expansions known from the mesonic sector
seemed to be lost \cite{Gasser:1987rb}, leading to the development of HBChPT \cite{Jenkins:1990jv,Bernard:1992qa}
in which one projects onto large and small components of the nucleon field and, finally, restores
a systematic counting scheme.
   This framework was successfully applied to many processes, including pion photo- and electroproduction
\cite{Bernard:1994gm,Bernard:1995cj,Bernard:2001gz,Bernard:1996ti,Bernard:1996bi,Bernard:1993bq}.
   However, giving up manifest Lorentz covariance may, in certain cases, lead to the wrong analytic behavior
of loop amplitudes \cite{Becher:1999he}.
   The seeming power-counting problem of RChPT for nucleons was addressed
using appropriate renormalization schemes such as, e.g., infrared regularization \cite{Becher:1999he} or the
extended on-mass-shell (EOMS) scheme \cite{Fuchs:2003qc}.

   The Lagrangian relevant to the one-nucleon sector consists of a purely mesonic part
($\mathcal{L}_{\pi}$) \cite{Gasser:1983yg} and a part containing the pion-nucleon interaction ($\mathcal{L}_{\pi N}$)
\cite{Gasser:1987rb,Fettes:2000gb},
\begin{equation}
\mathcal{L}=\mathcal{L}_{\pi}^{(2)}+\mathcal{L}_{\pi}^{(4)}
+\mathcal{L}_{\pi N}^{(1)}+\mathcal{L}_{\pi N}^{(2)}+\mathcal{L}_{\pi N}^{(3)}+\mathcal{L}_{\pi N}^{(4)}+\cdots.
\end{equation}
   The superscripts refer to the chiral order of the Lagrangians and the ellipsis stands for the neglected,
higher-order contributions.
   The Lagrangian contains a large number of LECs.
   Their numerical values cannot be derived from the effective field theory
itself but are determined by adjusting them to experimental data.
   Most of the LECs also enter simpler observables such as, e.g., form factors or the pion-nucleon coupling.
   In Table \ref{tab:lecs}, we display LECs of $\mathcal{L}_{\pi N}$ which have been extracted from processes
other than pion photoproduction.
\begin{table}[t]
            \caption{LECs determined from other processes.}
    \label{tab:lecs}
    \centering
        \begin{tabular}{ll}
        \hline
        \hline
    LEC       &     Source\\
    \hline
    $c_1$     &      proton mass $m_p=938.272$ MeV   \cite{Beringer:1900zz}        \\
    $c_2$, $c_3$, $c_4$ & pion-nucleon scattering \cite{Becher:2001hv}\\
    $c_6$, $c_7$ &  magnetic moment of proton ($\mu_p=2.793$) and neutron ($\mu_n=-1.913$) \cite{Beringer:1900zz}    \\
        $d_{16}$  &  axial-vector coupling constant $g_A=1.2695$  \cite{Beringer:1900zz}      \\
    $d_{18}$  &  pion-nucleon coupling constant $g_{\pi N}=13.21$ \cite{Schroder:2001rc}    \\
        \hline
        \hline
        \end{tabular}
\end{table}
   In the following, we focus on the contact interactions specific to pion photoproduction at
$O(q^3)$ and $O(q^4)$.
   The relevant part of the Lagrangian is given by
\begin{align}
\mathcal{L}^{(3)}_{\pi N}=&\frac{d_8}{2m_N}
\left[i\bar{\Psi}\epsilon^{\mu\nu\alpha\beta}\textnormal{Tr}\left(\tilde{f}_{\mu\nu}^+u_\alpha\right)D_\beta\Psi
+\textnormal{H.c.}\right]\nonumber\\
&+\frac{d_9}{2m_N}\left[i\bar{\Psi}\epsilon^{\mu\nu\alpha\beta}
\textnormal{Tr}\left(f_{\mu\nu}^++2v_{\mu\nu}^{(s)}\right)u_\alpha D_\beta\Psi
+\textnormal{H.c.}\right],\\
\mathcal{L}^{(4)}_{\pi N}=&-\frac{e_{48}}{4m_N}\left[i\bar{\Psi}\textnormal{Tr}\left(f_{\lambda\mu}^+
+2v_{\lambda\mu}^{(s)}\right)h^\lambda_\nu\gamma_5\gamma^\mu D^\nu\Psi+\textnormal{H.c.}\right]\nonumber\\
&-\frac{e_{49}}{4m_N}\left[i\bar{\Psi}\textnormal{Tr}\left(f_{\lambda\mu}^+ +2v_{\lambda\mu}^{(s)}\right)
h^\lambda_\nu\gamma_5\gamma^\nu D^\mu\Psi+\textnormal{H.c.}\right]\nonumber\\
&+\frac{e_{50}}{24m_N^3}\left[i\bar{\Psi}\textnormal{Tr}\left(f_{\lambda\mu}^+ +2v_{\lambda\mu}^{(s)}\right)
h_{\nu\rho}\gamma_5\gamma^\lambda D^{\mu\nu\rho}\Psi+\textnormal{H.c.}\right]\nonumber\\
&-\frac{e_{51}}{4m_N}\left[i\bar{\Psi}u^\lambda
[D_\lambda,\textnormal{Tr}\left(f_{\mu\nu}^+ +2v_{\mu\nu}^{(s)}\right)]\gamma_5\gamma^\mu D^\nu\Psi
+\textnormal{H.c.}\right]\nonumber\\
&-\frac{e_{67}}{4m_N}\left[i\bar{\Psi}\textnormal{Tr}\left(\tilde{f}_{\lambda\mu}^+h^\lambda_{\nu}\right)
\gamma_5\gamma^\mu D^\nu\Psi+\textnormal{H.c.}\right]\nonumber\\
&-\frac{e_{68}}{4m_N}\left[i\bar{\Psi} \textnormal{Tr}\left(\tilde{f}_{\lambda\mu}^+h^\lambda_{\nu}\right)
\gamma_5\gamma^\nu D^\mu\Psi+\textnormal{H.c.}\right]\nonumber\\
&+\frac{e_{69}}{24m_N^3}\left[i\bar{\Psi}\textnormal{Tr}\left(\tilde{f}_{\lambda\mu}^+h_{\nu\rho}\right)
\gamma_5\gamma^\lambda D^{\mu\nu\rho}\Psi+\textnormal{H.c.}\right]\nonumber\\
&-\frac{e_{71}}{4m_N}\left[i\bar{\Psi}\textnormal{Tr}\left(u^\lambda[D_\lambda,\tilde{f}_{\mu\nu}^+]\right)
\gamma_5\gamma^\mu D^{\nu}\Psi+\textnormal{H.c.}\right]\nonumber\\
&-\frac{e_{112}}{4m_N}\left[\bar{\Psi}\textnormal{Tr}\left(f_{\mu\nu}^++2v_{\mu\nu}^{(s)}\right)
\tilde{\chi}_-\gamma_5\gamma^\mu D^\nu\Psi+\textnormal{H.c.}\right]\nonumber\\
&-\frac{e_{113}}{4m_N}\left[\bar{\Psi}\textnormal{Tr}\left(\tilde{f}_{\mu\nu}^+\tilde{\chi}_-\right)
\gamma_5\gamma^\mu D^\nu\Psi+\textnormal{H.c.}\right],
\end{align}
where H.c.\ refers to the Hermitian conjugate.
   The nucleon is represented through the isospinor field $\Psi$, the pion appears after expanding
the so-called chiral vielbein $u_\mu$, and the photon is contained in the field-strength tensors
$f_{\mu\nu}^+$ and $v_{\mu\nu}^{(s)}$.
   For further definitions, the reader is referred to Ref.\ \cite{Fettes:2000gb}.
   In neutral pion photoproduction on the proton, only half of the LECs listed above can be determined,
because the (0) and (+) components of the isospin amplitudes [see Eq.\ (\ref{eqn:isospin})] both contribute
in the same way (see Appendix \ref{lecsandmult}).
   This reduces the number of independent LECs from twelve to six.
   In HBChPT, so far only five constants were considered, because that calculation took only $S$ and $P$ waves
into account.
   It turns out that at chiral order $O(q^4)$ another LEC appears which mainly affects
the multipole $E_{2-}$.
   In HBChPT one can rearrange the LECs such that two enter $E_{0+}$ and each of the three $P$ waves
comes with its own LEC.
   In the relativistic case the situation is more involved.
   A unique matching of the LECs to the multipoles cannot be done, which can be nicely seen in terms of the
$1/m_N$ expansion of the multipoles.
   At leading order, one reproduces the result of Ref.\ \cite{Bernard:2001gz}.
   At higher order, different linear combinations of the constants appear.

   Let us now address the renormalization condition.
   In the EOMS scheme, only terms explicitly violating the power counting are subtracted.
   From the six independent LECs of the contact diagrams only one is of $O(q^3)$.
   The other constants are of $O(q^4)$ and, therefore, are not necessary to subtract power-counting-violating
contributions, as we calculated the process up to and including $O(q^4)$.
   After a heavy-baryon expansion, the combination $d_8+d_9$ appears only in the multipole $P_3$
at order $O(q^3)$ (see Appendix \ref{lecsandmult}) and, therefore,
power-counting-violating terms appear only in this multipole.
   Furthermore, only diagrams of $O(q^4)$ can produce such contributions.
   The standard procedure to access the numerical value of such an LEC is through adjustment to
experimental data.
   Here, we exploit this fact to avoid the calculation of the power-counting-violating contribution
(see Appendix \ref{Renormalization} for further details).

   Another issue in pion production is isospin symmetry.
   As we work in the isospin-symmetric case, the cusp in the $E_{0+}$ multipole cannot appear.
   In Ref.\ \cite{Bernard:1993bq}, instead of the neutral-pion mass the mass of the $\pi^+$ was used in the loops,
leading to a phenomenologically correct description of the cusp.
   We also use this method to reproduce the cusp.
   Even though the effect is much smaller, we also use the neutron mass instead of the proton mass within the loops.
   The error one introduces this way is formally of higher order, because the mass difference of charged and
neutral pions and of proton and neutron is of higher order.

\section{Results and Discussion}
\subsection{The RChPT calculation}
   Up to and including chiral order four, 20 tree and 85 loop diagrams contribute to the more general case of
pion electroproduction.
   The topologies can, e.g., be found in Ref. \cite{Bernard:1992nc}.
   We calculated all diagrams with the aid of the computer algebra system Mathematica and the
FeynCalc package \cite{Mertig:1990an}.
   Even with a modern computer program such calculations are somewhat cumbersome.
   Furthermore, using computers always requires control over the programs used.
   In our case crossing symmetry and current conservation provide important checks on the results.
   All our results fulfill these requirements analytically.
   To evaluate loop integrals, we made use of the LoopTools package \cite{Hahn:2000kx}.

\begin{figure}[t]
    \centering
        \includegraphics[width=0.5\textwidth]{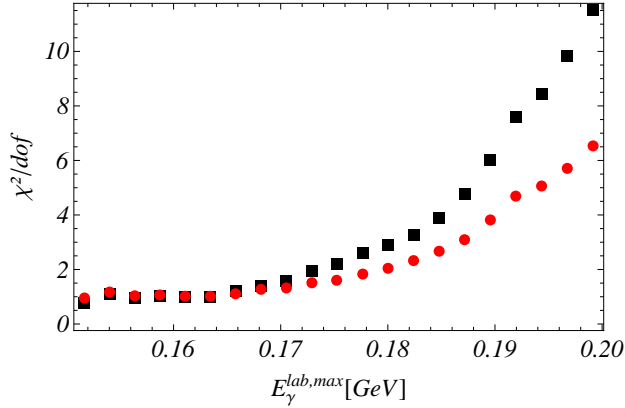}
    \caption{(color online) $\chi^2_{\textnormal{red}}$ as a function of the fitted energy range.
    The black squares and red dots refer to the RBChPT and HBChPT fits, respectively.}
    \label{fig:chi2plot}
\end{figure}

   Using the formulas for the multipole decomposition, we are able to project out any desired multipole.
   For our purposes we calculated $S$, $P$, and $D$ waves.
   The unknown LECs were determined via a $\chi^2$ fit to the latest MAMI data \cite{Hornidge:2012ca}.
   These were taken over a much wider energy range than ChPT can be applied to.
   Therefore, we had to determine the best energy range for a fit.
   The problem is that higher-energy data have smaller relative errors, leading to an increase of their
weight in a fit.
   We found $E_\gamma^\textnormal{lab}=165.8$ MeV to be a good maximum energy for the fit range.
   In Fig.\ \ref{fig:chi2plot}, we show how the $\chi^2_\textnormal{red}$ changes if one includes all
data points up to a certain energy $E_\gamma^\textnormal{lab,max}$.
   For comparison we also provide the reduced $\chi^2_\textnormal{red}$ of the HBChPT fit.

   It turned out that with the existing data we could only fit five of the six LECs.
   The problematic LEC, $\tilde{e}_{49}$, strongly influences the $E_{2-}$ multipole and, indirectly, also $E_{0+}$.
   Results for these multipoles, taking $\tilde{e}_{49}$ into account, are shown in Appendix \ref{e49problem}.
   In our final fit, we decided to set $\tilde{e}_{49}=0$ and obtained a minimal $\chi^2_\textnormal{red}$
   of 1.22.

   We estimate the errors of our parameters using the bootstrap method (see, e.g., Ref.\ \cite{efron}
for an introduction).
   Here, we only briefly outline the idea.
   If one has a data set $\mathbf{Y}=y_1,\ldots,y_n$ of length $n$, one can create $m$ bootstrap
samples $\mathbf{Y}_1,\ldots,\mathbf{Y}_m$ of length $n$, where $m$ should be a sufficiently large number.
   These new samples consist of the original data points, but randomly chosen.
   This means that in an arbitrary sample $\mathbf{Y}_k$ some data points appear  twice, three or even more times,
while others are neglected.
   Every sample can now be fitted in the same way as the original data.
   One ends up with $m$ values for the parameters.
   The idea behind the bootstrap is that the standard deviation for each parameter is an estimate of its error.
   Our results for the LECs including an error estimate are shown in Table \ref{tab:lecsppp}.
\begin{table}[t]
\caption{LECs of the contact diagrams. The $d_i$ are given in units of $\textnormal{GeV}^{-2}$ and
the $e_i$ in units of $\textnormal{GeV}^{-3}$. The errors stem from a bootstrap estimate (see text for details).}
    \label{tab:lecsppp}
    \centering
        \begin{tabular}{cc}
        \hline
        \hline
    LEC       &     Value\\
    \hline
$\tilde{d}_9:=d_8+d_9$          &  $-2.31\pm0.02$  \\
$\tilde{e}_{48}:=e_{48}+e_{67}$      &   $-3.0\pm0.2$     \\
$\tilde{e}_{49}:=e_{49}+e_{68}$      &   0    \\
$\tilde{e}_{50}:=e_{50}+e_{69}$      &   $-1.2\pm2.1$   \\
$\tilde{e}_{51}:=e_{51}+e_{71}$      &   $2.3\pm1.1$   \\
$\tilde{e}_{112}:=e_{112}+e_{113}$   &   $-4.4\pm2.1$   \\
        \hline
        \hline
        \end{tabular}
\end{table}

\begin{figure}[htbp]
    \centering
        \includegraphics[width=0.99\textwidth]{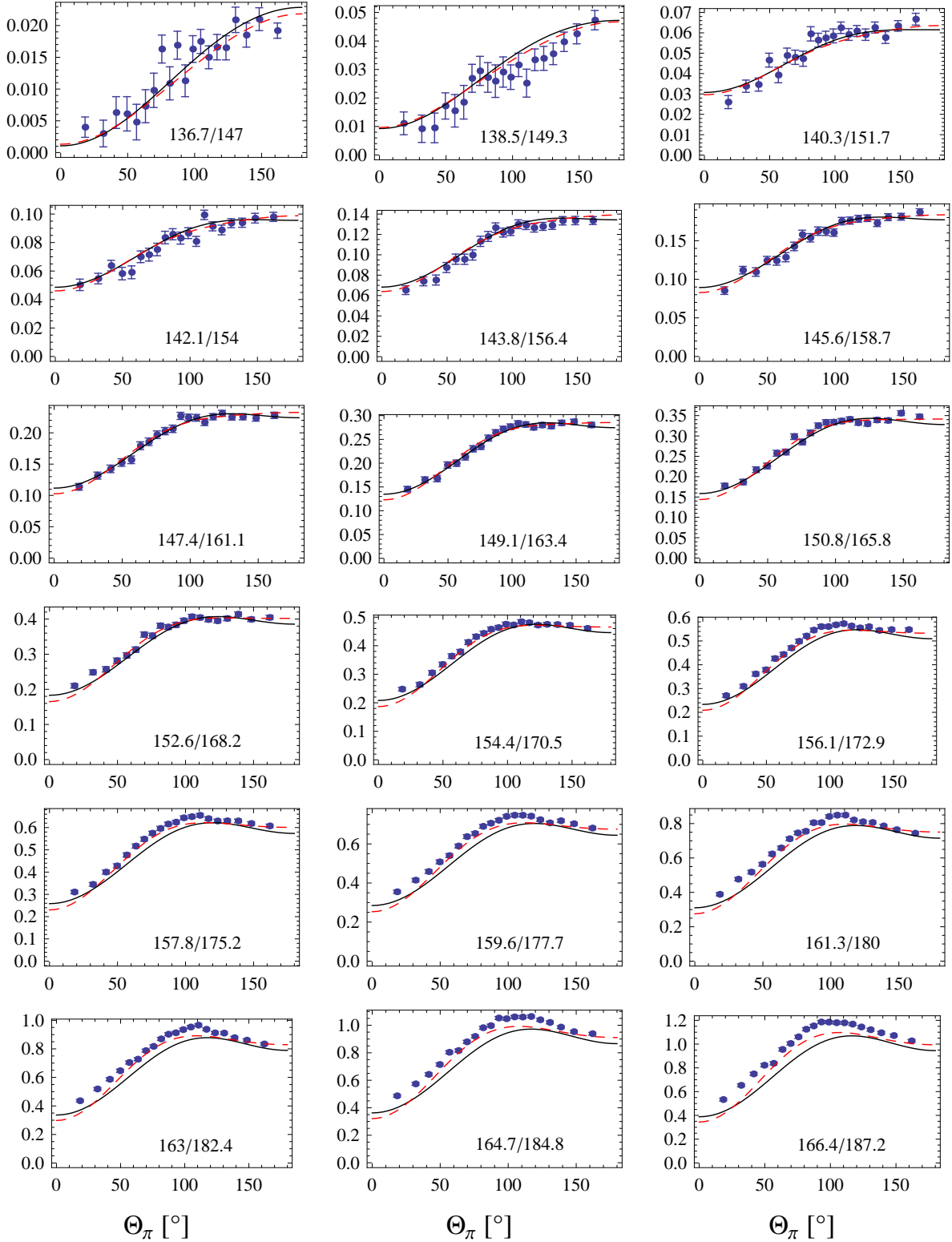}
    \caption{(color online)
    Differential cross sections $\sigma_0(\Theta_\pi)$ in $\mu b/sr$ as a function of the cm
    production angle $\Theta_\pi$.
    The graphs are shown for increasing pion energies in the cm frame/photon energies in the lab frame,
    both given in units of MeV.
    The solid (black) curves show the results in RChPT at $O(q^4)$, the dashed (red) curves show the
    same chiral order in HBChPT.
    The fits make use of data up to and including $E_\gamma^\textnormal{lab}=165.8$ MeV, i.e., the first
    nine figures.
    The data are taken from Ref.\ \cite{Hornidge:2012ca}.
    }
    \label{fig:wq}
\end{figure}

\begin{figure}[htbp]
    \centering
        \includegraphics[width=0.92\textwidth]{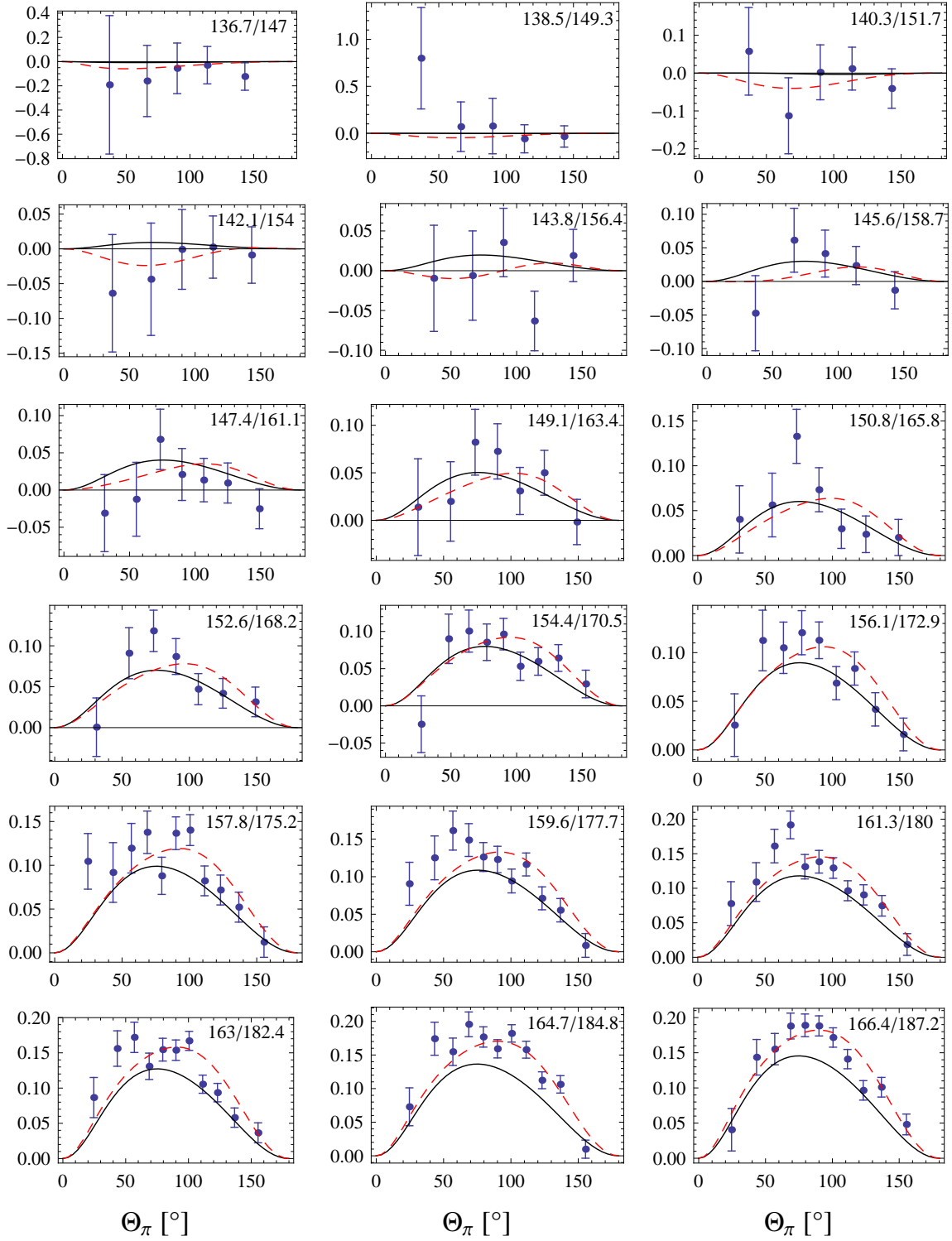}
    \caption{(color online) Photon asymmetries $\Sigma$ as a function of the cm production angle $\Theta_\pi$.
    The graphs are shown for increasing pion energies in the cm frame/photon energies in the lab frame,
    both given in units of MeV.
    The solid (black) curves show the results in RChPT at $O(q^4)$, the dashed (red) curves show the
    same chiral order in HBChPT.
    The fits make use of data up to and including $E_\gamma^\textnormal{lab}=165.8$ MeV, i.e., the first
    nine figures.
    The data are taken from Ref.\ \cite{Hornidge:2012ca}.
    }
    \label{fig:Sigma}
\end{figure}

\begin{figure}[htbp]
    \centering
        \includegraphics[width=\textwidth]{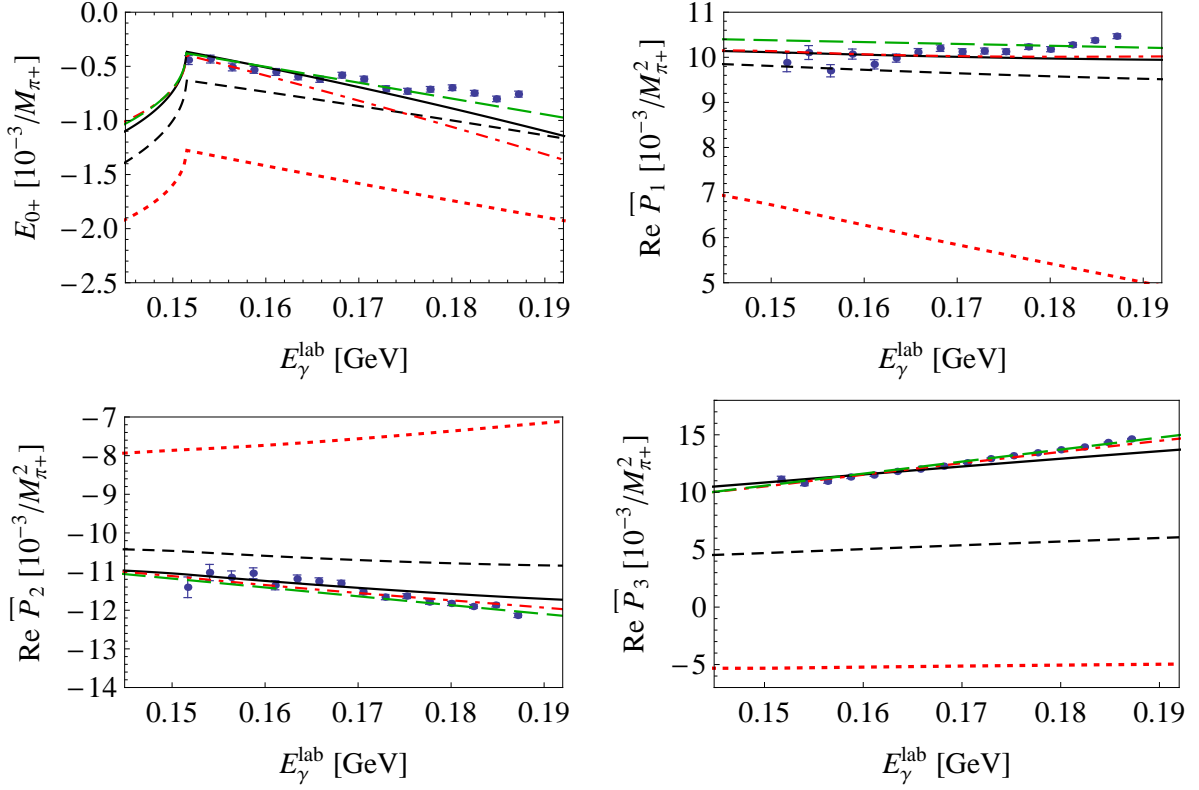}
    \caption{(color online) Real parts of the $S$- and $P$-wave multipoles as a function of $E_\gamma^\textnormal{lab}$.
    The solid (black) curves show the result in RChPT at $O(q^4)$, the dash-dotted (red) curves show the same
    chiral order in HBChPT.
    The dashed (black) lines show the RChPT result without the LECs, the dotted (red) lines show the HBChPT
    result without the LECs.
    The long-dashed (green) curves show the HBChPT fit of Ref.\ \cite{Hornidge:2012ca}.
    The data points are taken from a single-energy fit from Ref.\ \cite{Hornidge:2012ca}.}
    \label{fig:spwave}
\end{figure}

\begin{figure}[htbp]
    \centering
        \includegraphics[width=\textwidth]{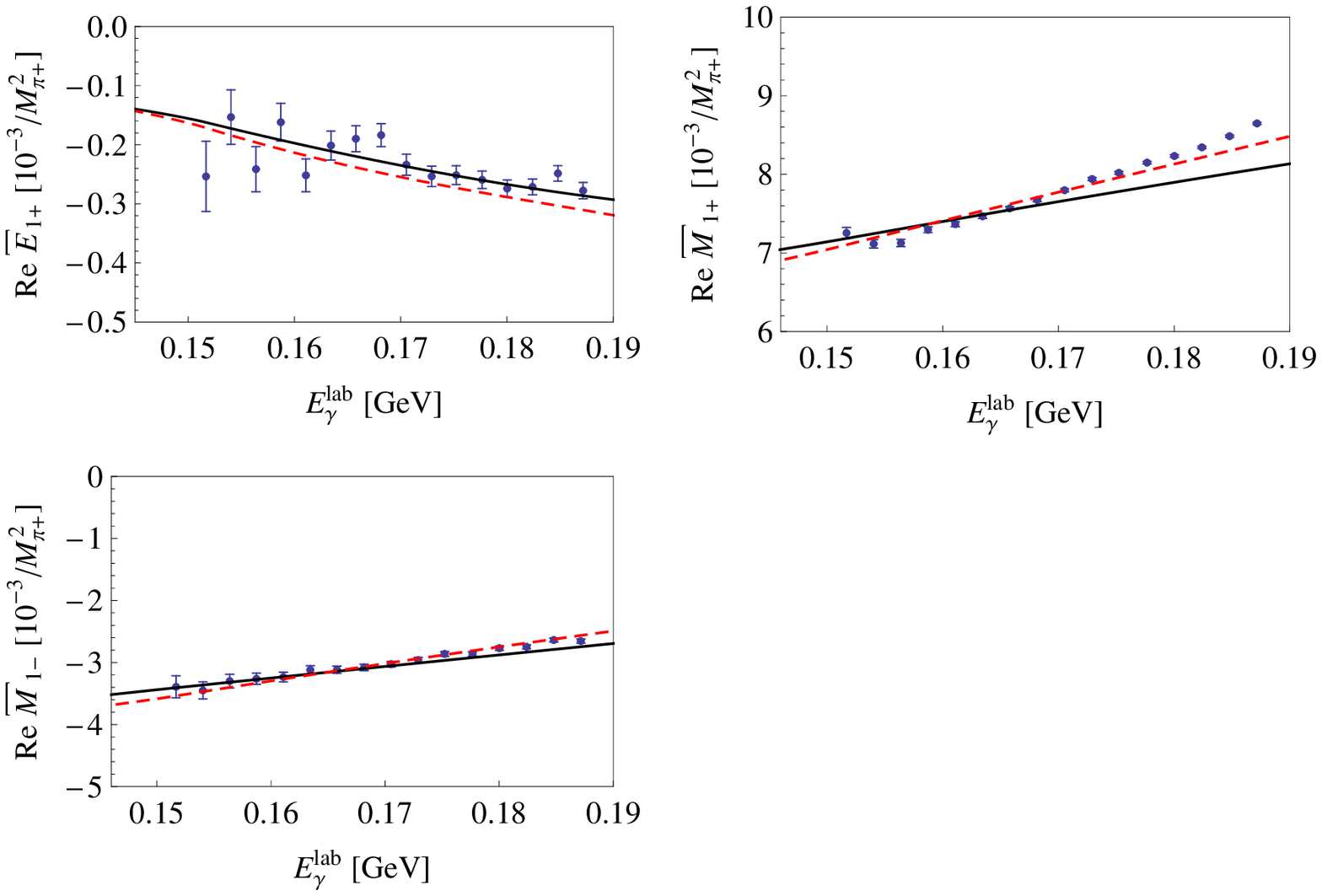}
    \caption{(color online) Physical $P$-wave multipoles as a function of $E_\gamma^\textnormal{lab}$.
    The solid (black) curves show the results in RChPT at $O(q^4)$, the dashed (red) curves show the
    same chiral order in HBChPT.
    The data points stem from a single-energy fit from Ref.\ \cite{Hornidge:2012ca}.}
    \label{fig:physpwaves}
\end{figure}

   The graphs for the measured differential cross sections and photon asymmetries are shown
in Figs.\ \ref{fig:wq} and \ref{fig:Sigma}.
   The corresponding multipoles are shown in Figs.\ \ref{fig:spwave} and \ref{fig:physpwaves}.
   The differential cross sections agree nicely with our result in the fitted energy range.
   For higher energies some differences between experiment and our calculation become visible.
   For the asymmetries the overall picture is similar; there, the difference at the highest energies
can be traced back to $P_2$ and $P_3$ which are a little bit too small compared to the single-energy fit
shown in Ref.\ \cite{Hornidge:2012ca}.
   The other multipoles agree up to approximately $E_\gamma^\textnormal{lab}=170$ MeV with the single-energy fits.
   In Fig.\ \ref{fig:physpwaves}, the physical $P$-wave multipoles are shown.
   Using this representation, one gets a clearer picture on the deviations from the data.
   The multipoles $E_{1+}$ and $M_{1-}$ agree nicely with the experiment.
   For the $M_{1+}$ one can see a rising of the data above $E_\gamma^\textnormal{lab}=170$ MeV,
which is related to the $\Delta$ resonance.
   As we did not include it explicitly, this rising does not appear in our curve.
   The most important $D$ wave, $E_{2-}$, is discussed in Appendix \ref{e49problem}.

\begin{figure}[htbp]
    \centering
        \includegraphics[width=0.633\textwidth]{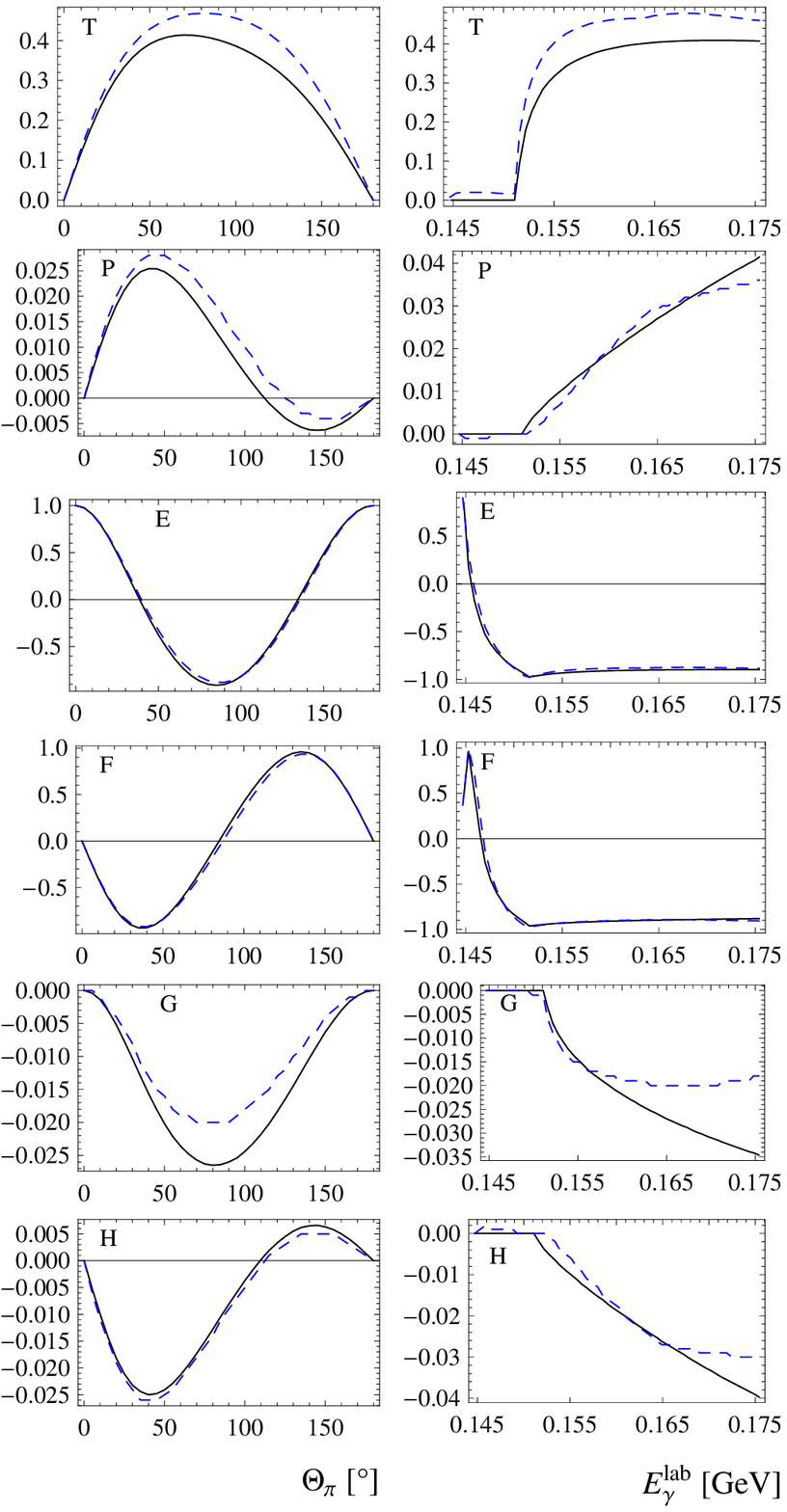}
    \caption{(color online)
    The polarization observables $T$, $P$, $E$, $F$, $G$, and $H$.
    The left column shows the angular dependence at $W=1090$ MeV and the right column shows the energy dependence
    of $T$, $E$, and $G$ for $\Theta_\pi=90^\circ$ and of $P$, $F$, and $H$ for $\Theta_\pi=45^\circ$, respectively.
    The solid (black) curves show the RChPT results and the dashed (blue) curves show the DMT model.
    }
    \label{fig:polobs}
\end{figure}

   As we now have the important multipoles in the threshold region, we can make some
predictions for upcoming experiments.
   Therefore, in Fig.\ \ref{fig:polobs} we show the polarization observables $T$, $P$, $E$, $F$, $G$, and $H$.
   Additionally, we show the predictions of the DMT model \cite{Kamalov:2000en}.
   We show the angular distribution at a fixed energy $W=1090$ MeV and the energy dependence at either
$\Theta_\pi=90^\circ$ or $\Theta_\pi=45^\circ$, depending on the approximate extreme value of
the observables.
   We find a good agreement between RChPT and DMT.

\subsection{Comparison with HBChPT}
   The HBChPT calculation of Ref.\ \cite{Bernard:2001gz} has been analyzed several times in the light of new
experimental data
\cite{Bernard:1995cj,Bernard:2001gz,FernandezRamirez:2009su,FernandezRamirez:2009jb,%
Schmidt:2001vg,FernandezRamirez:2012nw}.
   Here, we fit the LECs in the same way to the experiment as in the relativistic case,
allowing for a better comparison between both calculations.
   Note that the HBChPT fit of Ref.\ \cite{Hornidge:2012ca,FernandezRamirez:2012nw} extends to
a larger value of $E_\gamma^\textnormal{lab}$ which, as we have checked within our calculation,
partly explains the difference between the two HBChPT results in Fig.~\ref{fig:spwave}.
   Another source for the discrepancy is the use of different values for the coupling
constants and different fitting procedures.

   As mentioned before, $D$ waves are expected to be of some importance for extracting
the $E_{0+}$ multipole.
   We do not have the heavy-baryon result for the $D$ waves.
   Therefore, we use the Born terms to calculate $D$ waves.
   In addition, we included the heavy-baryon-expanded contribution of the sixth LEC to $E_{2-}$.
   Eventually, we experienced similar problems fitting this LEC as in the RChPT case.
   For that reason, we also set this constant to zero and used only the Born terms for the $D$ waves.
   With a value of 1.11, the $\chi^2_\textnormal{red}$ is better than in the relativistic case.
   One can see this in the observables too, as the heavy-baryon calculation seems to agree slightly
better with the data.
   The multipoles also support this picture.
   The LECs we obtained are listed in Table \ref{tab:lecshb}.
   On the other hand, the convergence properties of the relativistic result look more favorable.
   To illustrate this we also display the multipoles one gets, when switching off the LECs.
   The difference between the final result and this case is some indication of how good the series converges.
   Comparing the size of the LECs leads to the same conclusion.
   The most dramatic effect appears in the case of $P_3$, where in HBChPT the LEC completely
dominates the Born and loop terms.
   This gives us confidence that certain higher-order terms are very important here.
   The relativistic calculation keeps some corrections up to infinite order and this improves
the convergence behavior.

\begin{table}[tbp]
\caption{Values of the LECs in HBChPT obtained from a fit to Ref.\ \cite{Hornidge:2012ca}.
        We decided to set $e_{49HB}=0$ (see text). The errors stem from a bootstrap estimate (see text for details).  }
 \label{tab:lecshb}
    \centering
        \begin{tabular}{cc}
\hline
\hline
$a_1$      &  $(15.2\pm2.7)$      $\textnormal{GeV}^{-4}$\\
$a_2$      &  $(-7.6\pm2.5)$      $\textnormal{GeV}^{-4}$                  \\
$\xi_1$    &  $33.3\pm0.5$                        \\
$\xi_2$    &  $-31.8\pm0.5$                        \\
$b_p$      &  $(20.9\pm0.1)$      $\textnormal{GeV}^{-3}$                  \\
$e_{49HB}$&        0                  \\
\hline
\hline
\end{tabular}
\end{table}

\begin{figure}[htbp]
    \centering
        \includegraphics[width=0.5\textwidth]{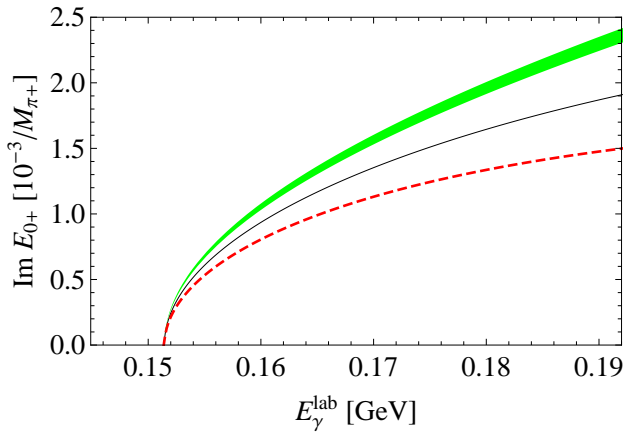}
    \caption{(color online)
    Imaginary part of $E_{0+}$ as a function of $E_\gamma^\textnormal{lab}$.
    The  solid (black) curve shows the result in RChPT at $O(q^4)$, the dashed (red) curve shows
    the same chiral order in HBChPT.
    The (green) band shows the result from unitarity with $\beta=(3.43\pm0.08)\times 10^{-3}/M_{\pi^+}$
    and $\gamma=0$.}
    \label{fig:ime0p}
\end{figure}

   Another interesting quantity one can derive from pion photoproduction is the so-called $\beta$ parameter
\cite{FernandezRamirez:2009jb} of the unitarity cusp \cite{Faldt:1979fs}.
   It is linked to pion-nucleon scattering and charged pion photoproduction via
\begin{equation}
\beta=M_{\pi^+}\,\textnormal{Re}[E_{0+}(\gamma,\pi^+)]\,a_\textnormal{cex}(\pi^+n\rightarrow\pi^0p)
=(3.43\pm0.08)\times 10^{-3}/M_{\pi^+},
\end{equation}
where the numerical estimate \cite{Hornidge:2012ca}
makes use of isospin symmetry to replace $a_\textnormal{cex}(\pi^+n\rightarrow\pi^0p)$
in terms of the experimentally known scattering length $a_\textnormal{cex}(\pi^-p\rightarrow\pi^0n)$.
   Close to threshold, unitarity connects this parameter to the imaginary part of $E_{0+}(\gamma,\pi^0)$,
\begin{equation}
\textnormal{Im}\left[E_{0+}(\gamma,\pi^0)\right]=\beta q_+,
\end{equation}
where $q_+$ is proportional to the three-momentum $|\vec{q}_+|$ of a $\pi^+$ in the cm frame,
\begin{equation}
q_+=|\vec{q}_+|/M_{\pi^+}.
\end{equation}
   To pin down the numerical value of $\beta$, we fit the imaginary part of $E_{0+}$ to the following series:
\begin{equation}
\textnormal{Im}\left[E_{0+}(\gamma,\pi^0)\right]=q_+\left(\beta +\gamma \frac{E_\gamma^\textnormal{lab}
-E_\gamma^\textnormal{lab,thr}}{M_{\pi^+}}\right).
\end{equation}
   In case of the relativistic calculation, we get $\beta_R=3.16\times 10^{-3}/M_{\pi^+}$ and
$\gamma_R=-1.08\times 10^{-3}/M_{\pi^+}$,
and HBChPT results in $\beta_{HB}=2.83\times 10^{-3}/M_{\pi^+}$ and
$\gamma_{HB}=-1.97\times 10^{-3}/M_{\pi^+}$.
   Both results are predictions, as all LECs were fixed in other processes,
including pion-nucleon scattering.
   Nevertheless, both results are too small compared to the experimental value of
$(3.43\pm0.08)\times 10^{-3}/M_{\pi^+}$ \cite{Hornidge:2012ca}.
   The relativistic result is somewhat closer, indicating that, again, certain higher-order contributions
are important.
   In Fig.\ \ref{fig:ime0p}, the imaginary part of $E_{0+}$ is shown as a function of $E_\gamma^\textnormal{lab}$.

\section{Summary}
   In this article we presented a full RChPT calculation up to $O(q^4)$ and one-loop order
for $\pi^0$ photoproduction on the proton.
   The amplitude was calculated with the aid of Mathematica and additional packages.
   Several tests were utilized to check our results.

   The next step was an analysis of the results in terms of multipoles.
   We took $S$, $P$, and $D$ waves into account, as these are the only relevant multipoles in the threshold region.
   The LECs of the contact diagrams were fitted using data of the latest MAMI experiment
for differential cross sections and photon asymmetries.
   Our calculation agrees well with the experimental data up to photon energies of approximately 170 MeV
   in the lab frame.

   We also discussed some of the properties of the LECs concerning the multipoles.
   We found that there are six LECs, of which one ($\tilde{e}_{49}$) is very important for
the determination of $E_{2-}$ and, therefore, also for $E_{0+}$.
   With the existing data we cannot narrow down this constant, because an unconstrained fit gives
an unnaturally large value for this LEC.
   Hence, we decided to neglect it in our main analysis.
   With the multipoles at hand we also gave some predictions for polarization observables for future experiments.

   In addition, we compared our results with HBChPT.
   Even though the latter seems to describe the experimental data slightly better,
the corresponding LECs have rather large values.
   This indicates that the relativistic calculation contains important higher-order effects.
   This can also be seen in the $\beta$ parameter, where the RChPT result is closer to the commonly
accepted value stemming from unitarity.

   Our numerical results are available through a web interface \cite{website}.
   The freedom to change the LECs gives one the opportunity to explore our results further in the
light of new upcoming data.
   We also give results for pion electroproduction and for all four physical reaction channels.
   The details will be discussed in a forthcoming publication \cite{pep}.

\begin{acknowledgements}
   This work was supported by the Deutsche Forschungsgemeinschaft (SFB 443 and 1044).
   The authors would like to thank D.~Drechsel, J.~Gegelia, and D.~Djukanovic for useful
discussions and support.
   We also thank the A2 and CB-TAPS collaborations for making available the experimental
data prior to publication.
\end{acknowledgements}

\begin{appendix}
\section{Heavy-baryon expansion of the contact contributions \label{lecsandmult}}
   Here, we provide the results of expanding the contact contributions to $E_{0+}$, the three $P$ waves,
and $E_{2-}$ in powers of $1/m_N$ up to and including next-to-leading order
(heavy-baryon expansion):
\begin{equation}
\mathcal{M}=\mathcal{M}^{[0]}+\frac{1}{m_N}\mathcal{M}^{[1]}+O\left(\frac{1}{m_N^2}\right).
\end{equation}
    Using the definitions of Table \ref{tab:lecsppp}, we obtain at leading order:
\begin{align}
E_{0+}^{[0]}&=\frac{e (6 \tilde{e}_{48}+2 \tilde{e}_{49}-4 \tilde{e}_{50}+3 \tilde{e}_{51})E_\pi^3}{12\pi F }
-\frac{e \left(3\tilde{e}_{112}+ \tilde{e}_{49} \right)M^2 E_\pi}{6  \pi  F},\\
\bar{P}_1^{[0]}&=-\frac{e (2 \tilde{e}_{48}+\tilde{e}_{51}) E_\pi^2}{4  \pi F},\\
\bar{P}_2^{[0]}&=\frac{e \tilde{e}_{48} E_\pi^2}{2 \pi F },\\
\bar{P}_3^{[0]}&=-\frac{e\tilde{d}_9 E_\pi}{\pi F},\\
\bar{\bar{E}}_{2-}^{[0]}&=-\frac{e \tilde{e}_{49} E_\pi}{6 \pi F}.\label{eqn:e2missue}
\end{align}
   At next-to-leading order, the results read
\begin{eqnarray}
E_{0+}^{[1]}&=&\frac{e (3 \tilde{e}_{112}+ \tilde{e}_{49}) M^4}{12 \pi F }
+\frac{e (6 \tilde{e}_{112}-12\tilde{e}_{48}-4\tilde{e}_{49}+12\tilde{e}_{50}
-5\tilde{e}_{51})M^2 E_\pi^2}{24 \pi F }\nonumber\\
&&+\frac{e (-3 \tilde{e}_{48}+\tilde{e}_{49}-2\tilde{e}_{50}-2\tilde{e}_{51})E_\pi^4}{12 \pi F },\\
\bar{P}_1^{[1]}&=&\frac{e (2 \tilde{e}_{48}-2\tilde{e}_{50}+\tilde{e}_{51}) E_\pi^3}{4 \pi F}
+\frac{e (-\tilde{e}_{112}+2\tilde{e}_{48}+\tilde{e}_{51}) M^2 E_\pi }{4 \pi F},\\
\bar{P}_2^{[1]}&=&\frac{e (6 \tilde{e}_{49}+4\tilde{e}_{50}-3\tilde{e}_{51}) E_\pi^3}{24 \pi F}
+\frac{e (\tilde{e}_{112}-2\tilde{e}_{48}) M^2 E_\pi}{4 \pi F},\\
\bar{P}_3^{[1]}&=&\frac{e \tilde{d}_9 M^2}{2 \pi F},\\
\bar{\bar{E}}_{2-}^{[1]}&=&\frac{e \tilde{e}_{49} M^2}{12 \pi F}
-\frac{e (6 \tilde{e}_{48}+4\tilde{e}_{49}+\tilde{e}_{51}) E_\pi^2}{48 \pi F}.
\end{eqnarray}
   As one can clearly see, only the lowest order allows for a rearrangement of
the LECs such that one can uniquely assign them to the different multipoles.
   For $P_3$ this also works up to and including first order.
   However, the other multipoles generate new mixings of the LECs at next-to-leading order.

\section{Renormalization of power-counting-violating contributions \label{Renormalization}}
   To some extent the EOMS scheme can be utilized to renormalize diagrams without explicitly
calculating the power-counting-violating part of the diagrams.
   Here, we explain this statement using a generic example, namely, the mass of a particle.
   Let us assume for the sake of simplicity that, after renormalization, power counting predicts a
tree-level contribution of chiral order $O(q^0)$ and a loop contribution of chiral order $O(q^2)$.
   Before renormalization, the mass is of the form
\begin{equation}
m=\textnormal{LEC}^0+\textnormal{Loop}^0,\label{eq:unrenmass}
\end{equation}
where $\textnormal{LEC}^0$ represents an unknown bare LEC and $\textnormal{Loop}^0$ represents
the unrenormalized loop contribution.
   In the following, we neglect any ultraviolet divergences in these expressions, i.e.,
we assume that they have been taken care of by applying the modified minimal subtraction
scheme of ChPT \cite{Gasser:1987rb}.
   We indicate this fact in terms of a superscript $r$.
   The infrared regular part of the loop contribution ($\textnormal{Loop}_{IR}^r$) can be
symbolically written as
\begin{equation}
\textnormal{Loop}^r_{IR}=\alpha_0^r+\alpha_2^r q^2+\alpha_4^r q^4+\cdots,\label{eq:unrenloop}
\end{equation}
where $q$ is a small quantity.
   For notational convenience, we have assumed only even powers of $q$.
   According to the above assumption, $\alpha_0^r$ violates the power counting.
   Renormalizing the loop contribution in the EOMS scheme amounts to subtracting the
power-counting-violating term $\alpha_0^r$ from $\textnormal{Loop}_{IR}^r$ \cite{Fuchs:2003qc}.
   In other words, $\textnormal{Loop}^r_{IR}$ is replaced by $\textnormal{Loop}^R_{IR}$,
and Eq.\ (\ref{eq:unrenloop}) simply becomes
\begin{equation}
\textnormal{Loop}^{R}_{IR}=\alpha_2^R q^2+\alpha_4^R q^4+\cdots.
\end{equation}
   Note that, in general, the higher-order coefficients $\alpha_2, \alpha_4, \ldots$ are
expressed in terms of EOMS-renormalized quantities.
   In this case, the expression for the mass reads
\begin{equation}
m=\textnormal{LEC}^R+\textnormal{Loop}^R,\label{eq:renmass}
\end{equation}
where the renormalized constant $\textnormal{LEC}^R$ now absorbs the
power-counting-violating part.
   Comparing Eqs.\ (\ref{eq:unrenmass}) and (\ref{eq:renmass}), the following connection can be made
\begin{equation}
\textnormal{LEC}^r+\alpha_0^r=\textnormal{LEC}^R.
\end{equation}
   The two expressions $\textnormal{LEC}^r$ and $\textnormal{LEC}^R$ only differ by the
(numerical) value of $\alpha_0^r$.
   In conclusion, adjusting LECs numerically is sufficient to renormalize diagrams.
   However, this procedure also has some drawbacks.
   First of all, only the sum of the adjusted LECs and the loop diagrams satisfies the power counting
whereas in the standard EOMS procedure each renormalized diagram satisfies the power
counting separately.
   As a consequence, in our calculation we cannot separate the fourth-order loop correction
for $P_3$.
   Another general problem is that the LEC does not necessarily have to be of natural size
anymore, as it contains a power-counting-violating part.
   In our specific case this is true for loop diagrams at fourth order that are renormalized
in terms of $\tilde{d}_9$.
   Nevertheless, in the present case, it turns out that this part is either small or has
the opposite sign and same magnitude of the numerical contribution of the renormalized coupling,
because even though $\tilde{d}_9$ contains a power-counting-violating part it is of natural size.

\section{The LEC $\tilde{e}_{49}$ \label{e49problem}}
\begin{figure}[htbp]
    \centering
        \includegraphics[width=\textwidth]{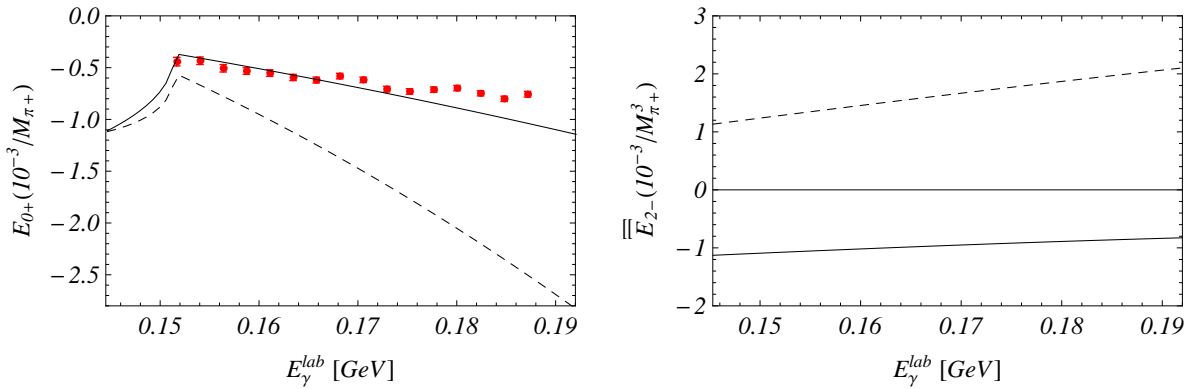}
    \caption{Results for the $E_{0+}$ and $E_{2-}$ multipoles
with $\tilde{e}_{49}=0$ (solid line) and $\tilde{e}_{49}=-35.7$
$\textnormal{GeV}^{-3}$ (dashed line). The latter value stems from a free fit to the data.}
    \label{fig:e49issue}
\end{figure}

   In principle, the LEC $\tilde{e}_{49}$ appears in all multipoles, but as one can see from
Eq.\ (\ref{eqn:e2missue}) it mainly affects $E_{2-}$.
   This multipole mixes strongly with $E_{0+}$.
   When performing a completely unconstrained fit, the solution with the lowest
$\chi^2_{\textnormal{red}}$ now is $\chi^2_{\textnormal{red}}=1.14$ and changes the values
of $E_{2-}$ significantly.
   The numerical value for $\tilde{e}_{49}$ becomes unnaturally large, as one can see from
Table \ref{tab:lece49}.
   The multipoles affected most are shown in Fig. \ref{fig:e49issue}.
   From the point of view of naturalness, it is very unlikely that an LEC picks up such a large value.
   Therefore, we believe such a large value to be an artifact until beam-target double
polarization observables $E$ and $F$ are available.
   This artifact can also be seen in the heavy-baryon calculation.
   There, we used the lowest-order contribution of the LEC for $E_{2-}$ [see Eq.\ (\ref{eqn:e2missue})],
as it would appear in exactly the same way in a true heavy-baryon calculation at $O(q^4)$
including $D$ waves.

\begin{table}[htbp]
    \centering
            \caption{LECs of the contact diagrams in a fit including $\tilde{e}_{49}$.
    The $d_i$ are given in units of $\textnormal{GeV}^{-2}$ and the $e_i$ in units of $\textnormal{GeV}^{-3}$.}
    \label{tab:lece49}
        \begin{tabular}{cc}
        \hline
        \hline
    LEC       &     Value\\
    \hline
$\tilde{d}_9:=d_8+d_9$          &  $-2.30$  \\
$\tilde{e}_{48}:=e_{48}+e_{67}$      &   $2.44$     \\
$\tilde{e}_{49}:=e_{49}+e_{68}$      &   $-35.7$    \\
$\tilde{e}_{50}:=e_{50}+e_{69}$      &   $1.8$   \\
$\tilde{e}_{51}:=e_{51}+e_{71}$      &   $-8.6$   \\
$\tilde{e}_{112}:=e_{112}+e_{113}$   &   $-4.8$   \\

        \hline
        \hline
        \end{tabular}
\end{table}

\end{appendix}

\end{document}